# The locality of quantum subsystems I

Adam Brownstein[1]

[1]*University of Melbourne, School of Physics*

23/08/2017

Many of the contemporary formulations of quantum mechanics describe the marginal probability distributions of entangled many-body systems in a non-local way. Unlike the non-locality of joint distributions, the non-locality of marginal distributions is not forced by theory or experiment. This paper investigates the issue in the context of the Copenhagen, de Broglie-Bohm and sum-over-paths interpretations. A dissociation between information flow into quantum subsystems and the tensor product structure of wavefunctions is highlighted in connection to the problem.

# 1 Introduction

Searching for an adequate ontology of physical processes can be a highly unconstrained endeavor, limited primarily by imagination[1]. Consequently, a scientific approach to ontology should attempt to constrain this search within the bounds of a set of reasonable guiding principles; for example by Occam's razor and local realism; at least to the maximum extent that experimental evidence will allow. Due to their prominence in scientific discourse, one might expect that the main interpretations of quantum mechanics are sufficiently constrained by such principles. However, this paper argues that these interpretations are not sufficiently constrained by local realism.

In probability theory there are two main types of probability to consider; joint probabilities and marginal probabilities. Joint probabilities describe the simultaneous measurement of multiple variables, and contain information about their correlations, while marginal probabilities describe the measurement outcomes of subsets of these variables in isolation. The two types of probability are connected through the law of total probability. Many-body quantum mechanics obeys the laws of probability theory, and therefore describes both types. In the quantum context, joint probabilities refer to the measurement outcomes for particles of the total system, while marginal probabilities refer to the measurement outcomes for particles of quantum subsystems; for instance in the two particle case, the measurement outcomes for particle A in the absence of knowledge of the outcomes for particle B.

Joint probabilities have received the majority of the theoretical attention to date. In the context of joint probabilities, Bell's theorem [1] has shown the impossibility of a local realist interpretation of the quantum predictions.[2] Therefore the question of whether the contemporary interpretations are sufficiently constrained by local realism may appear to lack relevance, since no interpretation can be local realist in a simple way. Nevertheless, it is perhaps neglected in this debate that quantum mechanics is also a theory of marginal probabilities.

In the context of quantum marginals, the question of local realism remains an open and interesting one. While entanglement correlations imply that the quantum mechanical 'total system' is more than a just product of its 'subsystems', these 'subsystems' may have a physical existence in their own right. It is quite plausible that the quantum marginal probabilities have a local realist explanation, which would enable subsystems to be described in an autonomous way. But as will be demonstrated in this paper, neither the Copenhagen, de Broglie-Bohm nor sum-over-paths interpretations provide a local realist explanation for the marginals. Consequently, these interpretations are not sufficiently constrained by local realism and may be vulnerable to the construction of alternative interpretations which have rectified this issue.

This paper investigates marginal probabilities in the Copenhagen, de Broglie-Bohm and sum-over-paths interpretations. Section two demonstrates that the Copenhagen interpretation of the marginals is non-local. The non-locality is observed to be a consequence of the breakdown in the interpretation of the marginal probabilities in terms of partial wavefunction collapse. Section three highlights that the issues encountered by the Copenhagen interpretation are shared by the de Broglie-Bohm interpretation of the marginals. Section four then formalizes what appears to be driving these difficulties, a disconnect between information flow into quantum subsystems and interactions encoded by the tensor product structure of the wavefunction. Section five shows that the standard sum-over-paths interpretation fails to rectify these issues of locality, and is furthermore susceptible to the violation of the hypotheses regarding information flow made in section four. The insights obtained in this paper serve as a foundation for papers two [12] and three [13] of this series.

---

[1] For instance as Bell (1982, [2]) remarked: "what is proved by impossibility proofs is lack of imagination.".

[2] Or more correctly, Bell's theorem has shown that novel ontology is required to achieve a local realist interpretation of quantum mechanics.

## 1.1 The configuration space problem

The issue of locality of the marginals is best understood in the context of a fundamental problem of the joint probabilities which has gained recognition in recent times. Since the wavefunction of an *N*-particle system contains a copy of the spatial coordinates of each constituent particle, a realist interpretation of the wavefunction seems to suggest that it inhabits a 3*N*-dimensional configuration space. It would more intuitive to describe all physical processes using a single set of spatial coordinates. But because the resultant probability density contains interference in this configuration space, the measurement results cannot be packaged into three-dimensional space in a simple way. We shall henceforth call this problem the *configuration space problem*. The configuration space problem has been described in the literature by Norsen [3], Lewis [4], Albert [5], Wallace & Timpson [6], Monton [7] and Allori [8] among others. Historically, the configuration space problem has been identified in particular by Einstein[3], de Broglie[4], and Bell[5]. For a brief sketch of the problem, consider the Schrödinger equation for two charged scalar particles:

$$i\hbar \frac{\partial \psi(\vec{x}_1, \vec{x}_2, t)}{\partial t} = \left[ -\frac{\hbar^2 \nabla_1^2}{2m} - \frac{\hbar^2 \nabla_2^2}{2m} + V_{E_1}(\vec{x}_1, t) + V_{E_2}(\vec{x}_2, t) + V_I(|\vec{x}_1 - \vec{x}_2|, t) \right] \psi(\vec{x}_1, \vec{x}_2, t), \quad (1.1)$$

where $V_{E_1}(\vec{x}_1, t)$ and $V_{E_2}(\vec{x}_2, t)$ are semiclassical potentials produced by the external system and experienced by particles 1 and 2 respectively, while $V_I(|\vec{x}_1 - \vec{x}_2|, t)$ is the Coulomb potential due to the particles' mutual interaction. Because the Coulomb potential $V_I(|\vec{x}_1 - \vec{x}_2|, t)$ depends on the coordinates of both particles, a wavefunction which is initially separable can become highly entangled. The non-separability $\psi(\vec{x}_1, \vec{x}_2, t) \neq \psi(\vec{x}_1, t)\psi(\vec{x}_2, t)$ then prevents the wavefunction from being split into two single-particle wavefunctions $\psi(\vec{x}_1, t)$ and $\psi(\vec{x}_2, t)$, which could otherwise be embedded into three-dimensional space. Consequently, the many-body Schrödinger equation (Eq. (1.1)) seems to inherently describe the motion of the wavefunction in a six-dimensional space configuration space. The reality of the configuration space description is evidenced by the probabilistic predictions of the configuration space wavefunction.

This series of papers addresses whether the marginal probabilities of entangled quantum systems share an analogous configuration space problem. It is shown that configuration space effects also appear in the marginal probabilities, and that these effects do not have an immediately apparent local interpretation. We denote this problem the *configuration space problem of quantum subsystems*. Finding a solution to this problem will require a new ontology for quantum subsystems.

---

[3] "Schrödinger is, in the beginning, very captivating. But the waves in n-dimensional coordinate space are indigestible" (Einstein, 1926 [9]).

[4] "Furthermore, if the propagation of a wave in space has a clear physical meaning, it is not the same as the propagation of a wave in the abstract configuration space, for which the number of dimensions is determined by the number of degrees of freedom of the system. " (de Broglie, 1927 [10]).

[5] "It is in the wavefunction that we must find an image of the physical world, and in particular of the arrangement of things in ordinary three-dimensional space. But the wavefunction as a whole lives in a much bigger space, of 3*N* dimensions." (Bell, 1987 [11]).

# 2 Marginals in the Copenhagen interpretation

Perhaps the most prevalent view on quantum mechanical marginal probabilities is that the no-signaling property ensures that the marginals are local. Once any many-body interactions have occurred and the constituent particles of the system are then spatially separated, the marginal probabilities are no longer influenced by the measurement of particles external to the isolated subsystems. It is only when the results of spatially separated measurements are brought together that non-local correlations can be observed.

However this prevalent view is quite misleading. In the case of quantum circuits displaying spatial entanglement, the marginal probabilities are difficult to reconcile with locality. The usual no-signaling arguments are stymied by the specificity of interactions between particles. To provide an example of this phenomenon, we study the eventual occurrence of configuration space effects in the marginals as a series of quantum gates are applied to a bipartite state. A breakdown in the local interpretation of these marginals can be observed in the Copenhagen description of partial wavefunction collapse.

## 2.1 First controlled phase gate

A single application of a controlled phase gate to a quantum circuit induces effects upon the marginal probabilities, but these effects have a simple local interpretation. Consider two qubits which become spatially entangled in the following quantum circuit. Suppose that Alice acts on particle A and Bob acts on particle B. Each qubit has two orthogonal spatial modes upon which the quantum information is encoded. The wavepackets comprising the spatial modes have disjoint spatial supports (i.e they are completely spatially separated with no overlapping non-zero regions) which prevents the propagation of information between them.

We utilize a quantum circuit notation that assigns a separate rail to each spatial mode. Most quantum circuits involve spin entangled qubits and are represented in a single rail notation. However, a single rail representation of spatial mode entangled qubits does not adequately emphasize the lack of communication between the spatial modes. This lack of communication is the source of the difficulties in the local interpretation of quantum marginals.

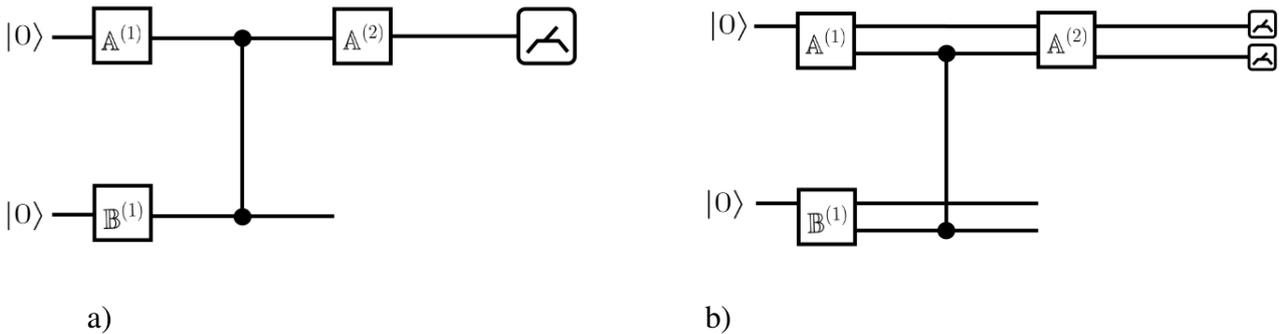

Figure 1: An initially separable state is prepared, and acted upon by a controlled phase gate. Alice acts upon the resultant state with a single particle gate before making a measurement. a) Represented in standard quantum circuit notation. b) Represented in a dual-rail notation to emphasize the local flow of information.

In the dual-rail representation of the circuit, the rails are connected to single particle unitary matrices displayed as labeled boxes. These unitary matrices serve to unitarily split up the input modes and distribute them amongst the output modes, and so can mix information obtained from both rails. A simple example of such a unitary operator is a Hadamard gate, which can be implemented physically using a 50-50 beamsplitter.

Controlled phase gates are represented as vertical rails which link up the spatial modes of two different particles. A simple example of a controlled phase gate can be found in the physical imple-mentation of photonic circuits, where the effect of Kerr optical nonlinearity induces a phase shift, conditional upon the whether a two-photon state is present in the nonlinear crystal. Controlled phase gates correspond to interactions occurring in a select region of configuration space, while single particle gates correspond to interactions occurring across extensive regions of configuration space.

The quantum circuit expresses the following operations on the quantum state. At step $t = 0$, the wavefunction is initialized into the $|0,0\rangle$ state. In matrix notation this state can be expressed as:

$$|\psi(0)\rangle = \begin{pmatrix} 1 & 0 & 0 & 0 \end{pmatrix}^T, \tag{2.1}$$

where we have used the vector basis: $\begin{pmatrix} 1 & 0 & 0 & 0 \end{pmatrix}^T \equiv |0,0\rangle$, $\begin{pmatrix} 0 & 1 & 0 & 0 \end{pmatrix}^T \equiv |0,1\rangle$, $\begin{pmatrix} 0 & 0 & 1 & 0 \end{pmatrix}^T \equiv |1,0\rangle$ and $\begin{pmatrix} 0 & 0 & 0 & 1 \end{pmatrix}^T \equiv |1,1\rangle$. At step $t = 1$, a set of single-particle gates $\mathbb{A}^{(1)}$ and $\mathbb{B}^{(1)}$ are applied to particles A and B, followed by a controlled phase gate $\mathbb{AB}^{(1)} = \text{diag}\{1, 1, 1, e^{i\theta_1}\}$. The state vector becomes:

$$\begin{aligned} |\psi(1)\rangle &= \mathbb{AB}^{(1)} \left( \mathbb{A}^{(1)} \otimes \mathbb{B}^{(1)} \right) |0,0\rangle \\ &= \begin{pmatrix} \alpha_0 \beta_0 & \alpha_0 \beta_1 & \alpha_1 \beta_0 & e^{i\theta_1} \alpha_1 \beta_1 \end{pmatrix}^T, \end{aligned} \tag{2.2}$$

where we have defined the matrix elements $A_{00}^{(1)} \equiv \alpha_0$, $A_{10}^{(1)} \equiv \alpha_1$ and $B_{00}^{(1)} \equiv \beta_0$, $B_{10}^{(1)} \equiv \beta_1$ for brevity. Refer to Appendix A for information on how a Kronecker product of matrices acts upon a state vector. If we now measure particle A, the resulting marginal probabilities have a straightforward local interpretation. The probability for observing particle A in mode $|0\rangle_A$ is:

$$\begin{aligned} P(A = 0) &= |\alpha_0|^2 |\beta_0|^2 + |\alpha_0|^2 |\beta_1|^2 \\ &= |\alpha_0|^2 \left( |\beta_0|^2 + |\beta_1|^2 \right) \\ &= |\alpha_0|^2, \end{aligned} \tag{2.3}$$

where we have used the unitary matrix identity $|\beta_0|^2 + |\beta_1|^2 = 1$. Now apply a second single-particle gate $\mathbb{A}^{(2)}$ to particle A. The state vector becomes:

$$|\psi(2)\rangle = \left( \mathbb{A}^{(2)} \otimes \mathbb{I} \right) \mathbb{AB}^{(1)} \left( \mathbb{A}^{(1)} \otimes \mathbb{B}^{(1)} \right) |0,0\rangle = \begin{pmatrix} \left( A_{00}^{(2)} \alpha_0 + A_{01}^{(2)} \alpha_1 \right) \beta_0 \\ \left( A_{00}^{(2)} \alpha_0 + A_{01}^{(2)} \alpha_1 e^{i\theta_1} \right) \beta_1 \\ \left( A_{10}^{(2)} \alpha_0 + A_{11}^{(2)} \alpha_1 \right) \beta_0 \\ \left( A_{10}^{(2)} \alpha_0 + A_{11}^{(2)} \alpha_1 e^{i\theta_1} \right) \beta_1 \end{pmatrix}. \tag{2.4}$$

If we measure particle A at this juncture, the marginal probabilites are seen to have a dependence on $|\beta_0|^2$ and $|\beta_1|^2$. For instance the marginal probability for measuring particle A in mode $|0\rangle_A$ is:

$$P(A = 0) = |A_{00}^{(2)} \alpha_0 + A_{01}^{(2)} \alpha_1|^2 |\beta_0|^2 + |A_{00}^{(2)} \alpha_0 + A_{01}^{(2)} \alpha_1 e^{i\theta_1}|^2 |\beta_1|^2. \tag{2.5}$$

We will now discuss the Copenhagen interpretation of this marginal probability. It is important to first define what we mean by 'Copenhagen interpretation', since there are many uses of the term. We regard the Copenhagen interpretation to be an interpretation of quantum mechanics where the system undergoes collapse (or partial collapse) upon measurement. This collapse might be understood in the popular sense of acquiring knowledge of the system, or in the more realist sense of an actual physical reduction of wavefunction. In both cases, the distinguishing feature of the Copenhagen interpretation appears to be that a) that the collapse process is irreversible, and b) the initial state collapses to a single subsequent state, which is weighted by the probability for collapse to this state. These features are important in distinguishing wavefunction collapse from local hidden-variable ontologies.

Now observe that the value $|\beta_0|^2$ is associated with the probability amplitude of the top rail of particle B, which is not in direct connection to particle A's subsystem. Therefore particle A cannot directly 'know' the value of $|\beta_0|^2$. However, we can take $|\beta_0|^2 = 1 - |\beta_1|^2$ to eliminate the dependence on $|\beta_0|^2$:

$$P(A=0) = |A^{(2)}_{00}\alpha_0 + A^{(2)}_{01}\alpha_1|^2 \left(1 - |\beta_1|^2\right) + |A^{(2)}_{00}\alpha_0 + A^{(2)}_{01}\alpha_1 e^{i\theta_1}|^2 |\beta_1|^2. \quad (2.6)$$

The marginal probability is now in the form of a local hidden variable theory, since $P(A=0) = P(A=0|\lambda_0)P(\lambda_0) + P(A=0|\lambda_1)P(\lambda_1)$, where $\lambda_0$ and $\lambda_1$ correspond to the events that particle B is present/absent from mode $|1\rangle_B$ during the application of the controlled gate respectively. To form this expression, we have taken $P(A=0|\lambda_0) = |A^{(2)}_{00}\alpha_0 + A^{(2)}_{01}\alpha_1|^2$, $P(\lambda_0) = \left(1 - |\beta_1|^2\right)$, $P(A=0|\lambda_1) = |A^{(2)}_{00}\alpha_0 + A^{(2)}_{01}\alpha_1 e^{i\theta_1}|^2$ and $P(\lambda_1) = |\beta_1|^2$. All of these terms can be locally determined by particle A.

### 2.1.1 What is a local hidden variable theory?

A hidden variable theory is an explanation for an observed physical quantity in terms of a collection of quantities which are unknown to the observer. It expresses the idea that the probability we observe may be the marginal probability of a larger system, which can be given formally as the condition:

$$P(A=a) = \sum_\lambda P(A=a,\lambda). \quad (2.7)$$

Assuming $P(A=a,\lambda) = P(A=a|\lambda)P(\lambda)$ this condition becomes:

$$P(A=a) = \sum_\lambda P(A=a|\lambda)P(\lambda). \quad (2.8)$$

A local hidden variable theory is a hidden variable theory where the joint probabilities $P(a,\lambda)$ are all locally obtainable for the production of $P(A=a)$. By comparing Eq. (2.8) to Eq. (2.6) the local hidden variable nature of the marginal probability in this example is evident.

### 2.1.2 Probability weights for partial collapse

The Copenhagen interpretation of this marginal probability (Eq. (2.6)) is to suggest that the states of particle A undergo a partial collapse to one of two possibilities. If particle B is observed by particle A to be in mode $|0\rangle_B$ during the application of the controlled phase gate, then the state of particle A partially collapses to something (i.e. a density matrix or a state of knowledge) which gives rise to the measurement probability $P(A=0) = |A^{(2)}_{00}\alpha_0 + A^{(2)}_{01}\alpha_1|^2$. Conversely, if particle B is observed by particle A to be in mode $|1\rangle_B$, then the state of particle A partially collapses to something which gives the measurement probability $P(A=0) = |A^{(2)}_{00}\alpha_0 + A^{(2)}_{01}\alpha_1 e^{i\theta_1}|^2$. The total probability for measuring $P(A=0)$ is a classical sum of these two probabilities, subsequently weighted by the probabilities for measuring the presence or absence of particle B respectively.

## 2.2 Second controlled phase gate

The problem with the Copenhagen interpretation is that the local interpretation of the marginals in terms of partial collapse becomes untenable in more complicated circuits. To witness a breakdown in the Copenhagen interpretation, apply another set of gates $\mathbb{B}^{(2)}$, $\mathbb{AB}^{(2)}$ and $\mathbb{A}^{(3)}$ to the circuit. The matrices $\mathbb{B}^{(2)}$ and $\mathbb{A}^{(3)}$ denote single-particle gates while the matrix $\mathbb{AB}^{(2)} = \mathrm{diag}\{1,1,1,e^{i\theta_2}\}$ denotes a second controlled phase gate.

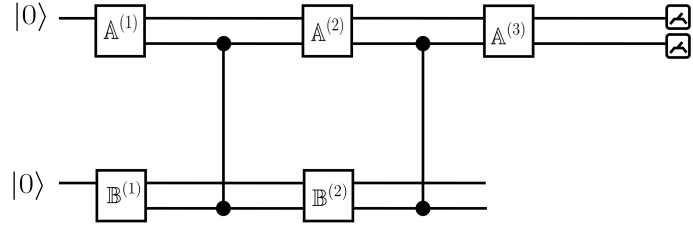

Figure 2: A separable state is prepared, then a sequence of single-particle and controlled phase gates are applied to the state. Alice acts a final single-particle gate upon the resultant state before making a measurement.

Recall that after the application of the gate $\mathbb{A}^{(2)}$, the wavefunction was:

$$|\psi(2)\rangle = \left(\mathbb{A}^{(2)} \otimes \mathbb{I}\right) \mathbb{AB}^{(1)} \left(\mathbb{A}^{(1)} \otimes \mathbb{B}^{(1)}\right) |0,0\rangle$$

$$= \begin{pmatrix} \left(A_{00}^{(2)}\alpha_0 + A_{01}^{(2)}\alpha_1\right)\beta_0 \\ \left(A_{00}^{(2)}\alpha_0 + A_{01}^{(2)}\alpha_1 e^{i\theta_1}\right)\beta_1 \\ \left(A_{10}^{(2)}\alpha_0 + A_{11}^{(2)}\alpha_1\right)\beta_0 \\ \left(A_{10}^{(2)}\alpha_0 + A_{11}^{(2)}\alpha_1 e^{i\theta_1}\right)\beta_1 \end{pmatrix}$$

$$= \begin{pmatrix} \alpha_0^{(2)}\beta_0 & \bar{\alpha}_0^{(2)}\beta_1 & \alpha_1^{(2)}\beta_0 & \bar{\alpha}_1^{(2)}\beta_1 e^{i\theta} \end{pmatrix}^T, \tag{2.9}$$

where for brevity, we have redefined the following terms:

$$\begin{pmatrix} A_{00}^{(2)}\alpha_0 + A_{01}^{(2)}\alpha_1 \end{pmatrix} \equiv \alpha_0^{(2)} \quad \begin{pmatrix} A_{00}^{(2)}\alpha_0 + A_{01}^{(2)}\alpha_1 e^{i\theta_1} \end{pmatrix} \equiv \bar{\alpha}_0^{(2)}$$
$$\begin{pmatrix} A_{10}^{(2)}\alpha_0 + A_{11}^{(2)}\alpha_1 \end{pmatrix} \equiv \alpha_1^{(2)} \quad \begin{pmatrix} A_{10}^{(2)}\alpha_0 + A_{11}^{(2)}\alpha_1 e^{i\theta_1} \end{pmatrix} \equiv \bar{\alpha}_1^{(2)}. \tag{2.10}$$

Apply the single-particle unitary $\mathbb{B}^{(2)}$ this state:

$$|\psi(2)\rangle = \left(\mathbb{A}^{(2)} \otimes \mathbb{B}^{(2)}\right) \mathbb{AB}^{(1)} \left(\mathbb{A}^{(1)} \otimes \mathbb{B}^{(1)}\right) |0,0\rangle$$

$$= \begin{pmatrix} B_{00}^{(2)}\alpha_0^{(2)}\beta_0 + B_{01}^{(2)}\bar{\alpha}_0^{(2)}\beta_1 \\ B_{10}^{(2)}\alpha_0^{(2)}\beta_0 + B_{11}^{(2)}\bar{\alpha}_0^{(2)}\beta_1 \\ B_{00}^{(2)}\alpha_1^{(2)}\beta_0 + B_{01}^{(2)}\bar{\alpha}_1^{(2)}\beta_1 \\ B_{10}^{(2)}\alpha_1^{(2)}\beta_0 + B_{11}^{(2)}\bar{\alpha}_1^{(2)}\beta_1 \end{pmatrix}$$

$$= \begin{pmatrix} m_1 & m_2 & m_3 & m_4 \end{pmatrix}^T, \tag{2.11}$$

where for brevity, we have redefined the following terms:

$$\begin{array}{ll} B_{00}^{(2)}\alpha_0^{(2)}\beta_0 + B_{01}^{(2)}\bar{\alpha}_0^{(2)}\beta_1 \equiv m_1 & B_{00}^{(2)}\alpha_1^{(2)}\beta_0 + B_{01}^{(2)}\bar{\alpha}_1^{(2)}\beta_1 \equiv m_3 \\ B_{10}^{(2)}\alpha_0^{(2)}\beta_0 + B_{11}^{(2)}\bar{\alpha}_0^{(2)}\beta_1 \equiv m_2 & B_{10}^{(2)}\alpha_1^{(2)}\beta_0 + B_{11}^{(2)}\bar{\alpha}_1^{(2)}\beta_1 \equiv m_4. \end{array} \tag{2.12}$$

In general, the constants $m_1, m_2, m_3, m_4 \in \mathbb{C}$ no longer have any common factors of matrix elements. It is at this point that the state vector is truly a configuration space concept. Apply another controlled phase gate $\mathbb{AB}^{(2)}$ to the state vector:

$$|\psi(2)\rangle = \mathbb{AB}^{(2)} \left(\mathbb{A}^{(2)} \otimes \mathbb{B}^{(2)}\right) \mathbb{AB}^{(1)} \left(\mathbb{A}^{(1)} \otimes \mathbb{B}^{(1)}\right) |0,0\rangle$$

$$= \begin{pmatrix} m_1 & m_2 & m_3 & m_4 e^{i\theta_2} \end{pmatrix}^T. \tag{2.13}$$

Finally, apply the single-particle gate $\mathbb{A}^{(3)}$ to particle A:

$$|\psi(3)\rangle = \left(\mathbb{A}^{(3)} \otimes \mathbb{I}\right) \mathbb{AB}^{(2)} \left(\mathbb{A}^{(2)} \otimes \mathbb{B}^{(2)}\right) \mathbb{AB}^{(1)} \left(\mathbb{A}^{(1)} \otimes \mathbb{B}^{(1)}\right) |0,0\rangle$$

$$= \begin{pmatrix} A_{00}^{(3)} m_1 + A_{01}^{(3)} m_3 \\ A_{00}^{(3)} m_2 + A_{01}^{(3)} m_4 e^{i\theta_2} \\ A_{10}^{(3)} m_1 + A_{11}^{(3)} m_3 \\ A_{10}^{(3)} m_2 + A_{11}^{(3)} m_4 e^{i\theta_2} \end{pmatrix}. \quad (2.14)$$

The marginal probabilities for Alice's subsystem are the following:

$$P(A=0) = P(A=0, B=0) + P(A=0, B=1)$$
$$= |A_{00}^{(3)} m_1 + A_{01}^{(3)} m_3|^2 + |A_{00}^{(3)} m_2 + A_{01}^{(3)} m_4 e^{i\theta_2}|^2. \quad (2.15)$$

$$P(A=1) = P(A=1, B=0) + P(A=1, B=1)$$
$$= |A_{10}^{(3)} m_1 + A_{11}^{(3)} m_3|^2 + |A_{10}^{(3)} m_2 + A_{11}^{(3)} m_4 e^{i\theta_2}|^2. \quad (2.16)$$

Examine the $|A_{00}^{(3)} m_1 + A_{01}^{(3)} m_3|^2$ term of $P(A=0)$ for instance:

$$|A_{00}^{(3)} m_1 + A_{01}^{(3)} m_3|^2 = |A_{00}^{(3)} \left(B_{00}^{(2)} \alpha_0^{(2)} \beta_0 + B_{01}^{(2)} \bar{\alpha}_0^{(2)} \beta_1\right) + A_{01}^{(3)} \left(B_{00}^{(2)} \alpha_1^{(2)} \beta_0 + B_{01}^{(2)} \bar{\alpha}_1^{(2)} \beta_1\right)|^2. \quad (2.17)$$

This term contains matrix elements for particle B which cannot be factorized outside the modulus. Previously the matrix elements for particle B were factorized outside the modulus and converted into amplitudes for measuring the presence or absence of particle B. We now have that $|A_{00}^{(3)} m_1 + A_{01}^{(3)} m_3|^2 \neq P(A=0|\lambda_0) P(\lambda_0)$ where $P(A=0|\lambda_0)$ and $P(\lambda_0)$ would be locally obtainable by particle A. Therefore the term $|A_{00}^{(3)} m_1 + A_{01}^{(3)} m_3|^2$ seems to be inherently produced by non-local wavefunction collapse. To convert Eq. (2.15) and (2.16) to a local hidden-variable form, it is necessary to perform further algebraic manipulations. However, it is not immediately obvious which algebraic manipulations should be applied. One approach would be to take:

$$P(A=0) = |A_{00}^{(3)} m_1 + A_{01}^{(3)} m_3|^2 + |A_{00}^{(3)} m_2 + A_{01}^{(3)} m_4 e^{i\theta_2}|^2$$
$$= |A_{00}^{(3)} m_1 + A_{01}^{(3)} m_3|^2 + |A_{00}^{(3)} m_2 + A_{01}^{(3)} m_4|^2 + 2Re\{A_{00}^{(3)*} A_{01}^{(3)} m_2^* m_4 \left(e^{i\theta_2} - 1\right)\}.$$
$$= P(A=0|\text{miss}) + 2Re\{A_{00}^{(3)*} A_{01}^{(3)} m_2^* m_4 \left(e^{i\theta_2} - 1\right)\}, \quad (2.18)$$

where we have identified $P(A=0|\text{miss}) = |A_{00}^{(3)} m_1 + A_{01}^{(3)} m_3|^2 + |A_{00}^{(3)} m_2 + A_{01}^{(3)} m_4|^2$ as the marginal probability for particle A given that the $\mathbb{AB}^{(2)}$ gate is removed from the circuit. A no-signaling condition can be used to remove the dependencies upon particle B from $P(A=0|\text{miss})$. But the second term, $2Re\{A_{00}^{(3)*} A_{01}^{(3)} m_2^* m_4 \left(e^{i\theta_2} - 1\right)\}$ is problematic, since it is a complex number in general. Therefore this second term does not represent partial wavefunction collapse. Consequently, the two terms of equation 2.18 cannot be understood in a Copenhagen framework.[6] Furthermore, the interpretation of the second term is difficult, since it depends on $m_2$, which is the probability amplitude for non-local wavefunction collapse to the state $|01\rangle$ just prior to the interaction.

---

[6]These two terms are also pieces of a reduced density matrix ontology. However the question of specifying an ontology (reduced density matrices or otherwise) is secondary. The corresponding reduced density matrix ontology for these terms cannot be understood in the traditional Copenhagen sense of partial wavefunction collapse.

## 2.3 Summary of results

This section has examined the gradual occurrence of configuration space effects in marginal probability distributions of a simple quantum circuit. The key message is that after applying a sequence of controlled phase gates, the state vector acquires the form of $|\psi\rangle = \begin{pmatrix} m_1 & m_2 & m_3 & m_4 \end{pmatrix}^T$, where $m_1, m_2, m_3, m_4 \in \mathbb{C}$ have no common factors of matrix elements of the circuit. It is at this point that the wavefunction is highly entangled, and should be regarded as a configuration space concept.

When the wavefunction is of this form, the marginal probabilities loose their ability to be locally interpreted in terms of partial wavefunction collapse to probability weighted states. Because we regard collapse to probability weighted states as the central feature of the Copenhagen interpretation, we claim that the Copenhagen interpretation of the marginal probabilities is non-local.

# 3 Marginals in the de Broglie-Bohm interpretation

This section investigates the locality of marginal probabilities in the de Broglie-Bohm interpretation. First, we discuss the concept of the conditional wavefunction, which is the main method of dealing with subsystems in the de Broglie-Bohm interpretation. The conditional wavefunction approach is evidently non-local. Second, it is shown that an initially more promising way to construct a de Broglie-Bohm interpretation of the subsystem suffers from similar issues to the Copenhagen interpretation of the subsystem.

## 3.1 The conditional wavefunction

The main way to describe quantum subsystems in the de Broglie-Bohm interpretation is through the conditional wavefunction. The conditional wavefunction is the total wavefunction conditioned upon the coordinates of particles external to the subsystem. Define $\vec{x}$ and $\vec{y}$ to be tuples in the configuration spaces pertaining to particles of the subsystem and external system respectively. Similarly, define $\vec{X}(t)$ are $\vec{Y}(t)$ to be tuples specifying the locations of the actual particle configurations for the subsystem and external system respectively, at time $t$. Then the following conditional wavefunction Eq. (3.1) provides a notion of wavefunction for the subsystem.

$$\Psi(\vec{x}, \vec{Y}(t), t) = \Psi(\vec{x}, \vec{y}, t)|_{\vec{y}=\vec{Y}(t)}. \tag{3.1}$$

The conditional wavefunction has a non-local dependence on the external system through the conditioning procedure $|_{\vec{y}=\vec{Y}(t)}$. In special cases, the effect of this non-locality disappears at the level of the de Broglie-Bohm guidance equations, leading to the concept of an *effective wavefunction*. Mathematically, if $\psi_1(\vec{x},t), \psi_2(\vec{y},t)$ and $\psi_3(\vec{x},\vec{y})$ are arbitrary wavefunctions and:

$$\begin{aligned}\Psi(\vec{x}, \vec{Y}(t), t) &= \Psi(\vec{x}, \vec{y}, t)|_{\vec{y}=\tilde{\vec{Y}}(t)} \\ &= [\psi_1(\vec{x})\psi_2(\vec{y}) + \psi_3(\vec{x},\vec{y})]|_{\vec{y}=\vec{Y}(t)} \\ &= \psi_1(\vec{x})\psi_2(\vec{Y}(t)) + \psi_3(\vec{x},\vec{Y}(t)) \\ &= \psi_1(\vec{x})\psi_2(\vec{Y}(t)), \end{aligned} \tag{3.2}$$

where $\psi_3(\vec{x},\vec{Y}(t)) \approx 0$, then $\psi_1(\vec{x})$ is an effective wavefunction for the subsystem. The effective wavefunction undergoes a Schrödinger time evolution in the configuration space of the subsystem, to produce the correct guidance equations for the subsystem configuration $\vec{X}(t)$. Nevertheless, even in the cases where it can be defined, the effective wavefunction is not sufficient to provide a local interpretation of the subsystem. There remains a lack of explanation for the transition between the conditional wavefunction description and the effective wavefunction description. For instance the particle configuration $\vec{Y}(t)$ can enter a region which is disjoint from the spatial support of $\psi_3(\vec{x},\vec{y},t)$, causing the conditional wavefunction to become an effective wavefunction. However the particle configuration $\vec{X}(t)$ cannot determine that this event has occurred without non-local communication.

## 3.2 A de Broglie interpretation of the subsystem

A more promising way to provide a local de Broglie-Bohm interpretation for the subsystem is to construct a separate ontology for the subsystem directly from the continuity equation. This allows the explicit non-locality of the conditional wavefunction description to be avoided. Consider the continuity equation for the configuration space probability density:

$$\frac{\partial |\psi(\vec{x}_1,...,\vec{x}_n,t)|^2}{\partial t} + \sum_i \nabla_i \cdot \mathbf{j}_i(\vec{x}_1,...,\vec{x}_n,t) = 0. \tag{3.3}$$

Integrate this probability density and examine the continuity equation for the marginal probability density of particle 1.

$$0 = \frac{\partial \int |\psi(\vec{x}_1,...,\vec{x}_n,t)|^2 d\vec{x}_2...d\vec{x}_n}{\partial t} + \sum_i \nabla_i \cdot \left( \int \mathbf{j}_1(\vec{x}_1,...,\vec{x}_n,t) d\vec{x}_2...d\vec{x}_n \right)$$

$$= \frac{\partial |\psi(\vec{x}_1,t)|^2}{\partial t} + \nabla_1 \cdot \left( |\psi(\vec{x}_1,t)|^2 \frac{\mathbf{j}_1(\vec{x}_1,t)}{|\psi(\vec{x}_1,t)|^2} \right),$$

where we have used the divergence theorem to set $\int \nabla_{i \neq 1} \cdot (\mathbf{j}_1(\vec{x}_1,...,\vec{x}_n,t)) d\vec{x}_2...d\vec{x}_n = 0$ each individually to zero, and have taken $|\psi(\vec{x}_1,t)|^2 = \int |\psi(\vec{x}_1,...,\vec{x}_n,t)|^2 d\vec{x}_2...d\vec{x}_n$ and $\mathbf{j}_1(\vec{x}_1,t) = \int \mathbf{j}_1(\vec{x}_1,...,\vec{x}_n,t) d\vec{x}_2...d\vec{x}_n$. To remain equivariant with the marginal probability density, the de Broglie-Bohm particles of the subsystem must have the following guidance equation:

$$\frac{d\mathbf{X}_1(t)}{dt} = \left. \frac{\mathbf{j}_1(\vec{x}_1,t)}{|\psi(\vec{x}_1,t)|^2} \right|_{\mathbf{X}_1=\vec{x}_1} = \left. \frac{\int \mathbf{j}_1(\vec{x}_1,...,\vec{x}_n,t) d\vec{x}_2...d\vec{x}_n}{\int |\psi(\vec{x}_1,...,\vec{x}_n,t)|^2 d\vec{x}_2...d\vec{x}_n} \right|_{\mathbf{X}_1=\vec{x}_1}. \tag{3.4}$$

The de Broglie-Bohm interpretation of the subsystem given by Eq. (3.4) has been investigated by Luis & Sanz [14] and Nikolić [15]. The problem with this guidance equation is that it depends on the joint probability $|\psi(\vec{x}_1,...,\vec{x}_n,t)|^2$ and configuration space current $\mathbf{j}_1(\vec{x}_1,...,\vec{x}_n,t)$ through $\int |\psi(\vec{x}_1,...,\vec{x}_n,t)|^2 d\vec{x}_2...d\vec{x}_n$ and $\int \mathbf{j}_1(\vec{x}_1,...,\vec{x}_n,t) d\vec{x}_2...d\vec{x}_n$ respectively. Unless these two quantities both have a local interpretation, the corresponding de Broglie-Bohm interpretation of the subsystem will be non-local.

## 3.3 Summary of results

Conditional wavefunctions in the de Broglie-Bohm interpretation describe quantum subsystems in a non-local way. To go beyond the conditional wavefunction description, guidance equations for the subsystem can be derived by integrating the continuity equation. However, this second approach retains an implicit dependence on the configuration space probability density and current.

# 4 Information flow in quantum subsystems

In this section, we lay out a set of intuitive hypotheses regarding the nature of information flow in quantum subsystems. It is subsequently shown that quantum marginals violate these hypotheses. This provides a strong constraint on the types of local hidden-variable theories which can reproduce quantum marginals.

### 4.0.1 Hypotheses regarding information flow in quantum subsystems:

The two hypotheses about information flow in quantum subsystems are the following:

1. *Information cannot enter the subsystem unless there is a physical medium carrying this information.*

   This hypothesis implies that the standard quantum mechanical description of non-local updating of the many-body wavefunction cannot be used to propagate information into the subsystem. Furthermore, it is safe to assume that information cannot flow between spatial modes that have disjoint spatial supports, because in principle, it is possible to separate these spatial modes by large distances before performing any interactions between them.

2. *The information which enters the subsystem is solely that which is obtained from the components of the configuration space wavefunction undergoing the interaction.*

   This hypothesis suggests the only way for updates of the many-body wavefunction to affect the subsystem is through the transmission of information directly via interactions described by standard quantum mechanics. Quantum mechanical interactions always occur in accordance with the tensor product structure of the wavefunction. For an illustration of this hypothesis, consider a two qubit system entangled in spatial mode. Suppose that the wavefunction is in the state:
   $$|\psi\rangle = \begin{pmatrix} m_1 & m_2 & m_3 & m_4 \end{pmatrix}^T,$$
   where $m_i \in \mathbb{C}$ are arbitrary complex coefficients. This hypothesis means, for example, that if modes $|1\rangle_A$ and $|1\rangle_B$ of particles A and B undergo a two body interaction, the only information which can enter particle A's subsystem is contained in the coefficient $m_4$.

### 4.0.2 Demonstration of violation of the hypothesis:

Suppose we prepare the following Bell state $|\psi(0)\rangle$ entangled in spatial mode:

$$|\psi(0)\rangle = \frac{1}{\sqrt{2}} \begin{pmatrix} 1 & 0 & 0 & 1 \end{pmatrix}^T. \tag{4.1}$$

Apply a single particle gate $\mathbb{B}$ to particle B, where $\mathbb{B}$ is:

$$\mathbb{B} = \begin{pmatrix} B_{00} & B_{01} \\ B_{10} & B_{11} \end{pmatrix}. \tag{4.2}$$

The wavefunction after this gate is:

$$|\psi(1)\rangle = (\mathbb{I} \otimes \mathbb{B})|\psi(0)\rangle$$
$$= \frac{1}{\sqrt{2}} \begin{pmatrix} B_{00} & B_{10} & B_{01} & B_{11} \end{pmatrix}^T. \tag{4.3}$$

By hypothesis one, Alice's subsystem cannot yet know any of the matrix elements $B_{00}$, $B_{01}$, $B_{10}$ or $B_{11}$ since these may be are remotely chosen by Bob. Now apply a controlled phase gate $\mathbb{AB}^{(1)}$, where $\mathbb{AB}^{(1)} = \text{diag}\left(1, 1, 1, e^{i\theta_1}\right)$. The wavefunction becomes:

$$|\psi(1)\rangle = \mathbb{AB}^{(1)} (\mathbb{I} \otimes \mathbb{B}) |\psi(0)\rangle = \frac{1}{\sqrt{2}} \begin{pmatrix} B_{00} & B_{10} & B_{01} & B_{11} e^{i\theta_1} \end{pmatrix}^T. \quad (4.4)$$

By hypothesis two, particle A is able to extract information from the coefficient $e^{i\theta_1} B_{11}$ during this interaction. Then a Hadamard gate is applied to particle A:

$$|\psi(2)\rangle = (\mathbb{H} \otimes \mathbb{I}) \mathbb{AB}^{(1)} (\mathbb{I} \otimes \mathbb{B}) |\psi(0)\rangle = \frac{1}{2} \begin{pmatrix} B_{00} + B_{01} \\ B_{10} + B_{11} e^{i\theta_1} \\ B_{00} - B_{01} \\ B_{10} - B_{11} e^{i\theta_1} \end{pmatrix}. \quad (4.5)$$

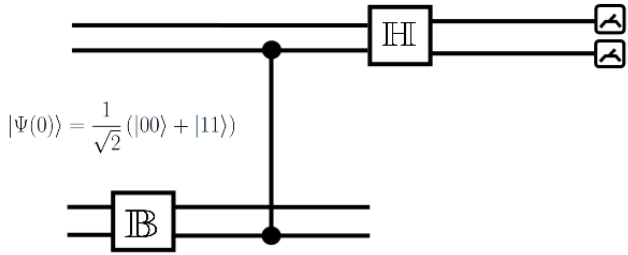

Figure 3: A Bell state $|\Psi(0)\rangle$ is prepared and undergoes the quantum circuit depicted. Alice makes a measurement and observes a sample from the marginal probability distribution of particle A.

At this point, Alice performs a measurement upon her subsystem. The marginal probabilities for particle A are:

$$P(A=0) = P(A=0, B=0) + P(A=0, B=1)$$
$$= \frac{1}{4} |B_{00} + B_{01}|^2 + \frac{1}{4} |B_{10} + B_{11} e^{i\theta_1}|^2$$
$$= \frac{1}{2} \left[1 + \text{Re}\{B_{10}^* B_{11} \left(e^{i\theta_1} - 1\right)\}\right].$$

$$P(A=1) = P(A=1, B=0) + P(A=1, B=1)$$
$$= \frac{1}{4} |B_{00} - B_{01}|^2 + \frac{1}{4} |B_{10} - B_{11} e^{i\theta_1}|^2$$
$$= \frac{1}{2} \left[1 - \text{Re}\{B_{10}^* B_{11} \left(e^{i\theta_1} - 1\right)\}\right].$$

Where we have expanded out the squared modulus and used the following unitary matrix identities $|B_{00}|^2 + |B_{01}|^2 = 1$, $|B_{10}|^2 + |B_{11}|^2 = 1$ and $B_{00}^* B_{01} + B_{10}^* B_{11} = 0$ to perform cancellations. Clearly particle A's marginal probability depends on $B_{10}^*$, which contradicts hypothesis two. For instance we can take the matrix $\mathbb{B}$ to be:

$$\mathbb{B} = \frac{1}{\sqrt{2}} \begin{pmatrix} e^{i\beta_1} & e^{i\beta_2} \\ e^{-i\beta_2} & -e^{-i\beta_1} \end{pmatrix}, \quad (4.6)$$

where $e^{i\beta_1}$ and $e^{i\beta_2}$ are arbitrary phases. Then particle A's marginal probabilities become:

$$P(A=0) = \frac{1}{2} \left[1 + \text{Re}\{B_{10}^* B_{11} \left(e^{i\theta_1} - 1\right)\}\right] = \frac{1}{2} \left[1 - \text{Re}\{e^{i\beta_2} e^{-i\beta_1} \left(e^{i\theta_1} - 1\right)\}\right], \quad (4.7)$$

$$P(A=1) = \frac{1}{2} \left[1 - \text{Re}\{B_{10}^* B_{11} \left(e^{i\theta_1} - 1\right)\}\right] = \frac{1}{2} \left[1 + \text{Re}\{e^{i\beta_2} e^{-i\beta_1} \left(e^{i\theta_1} - 1\right)\}\right]. \quad (4.8)$$

Given knowledge of the parameters $e^{i\beta_1}$ and $e^{i\theta_1}$, Alice can use her measurement results to deduce information about the parameter $e^{i\beta_2}$. But by hypothesis 2, Alice should not be able to deduce this information, since this phase $e^{i\beta_2}$ was not associated with the part of the configuration space wavefunction that undergoes the two particle interaction. What actually appears to be occurring is that Alice's part of the state $|1\rangle_A|1\rangle_B$ is gleaning information from Bob's part of the state $|0\rangle_A|1\rangle_B$. There is an evident dissociation between the interactions defined by the tensor product structure of the wavefunction and the propagation of information into the subsystem.

Figure 4 provides a visualization of configuration space wavepackets embedded into three-dimensional space in a quantum circuit to illustrate this disconnection. The wavepackets are assumed to communicate only to their partners as defined by the tensor product structure of the wavefunction. This section has shown to the contrary that the tensor product structure cannot be used to label wavepackets as communicating partners.

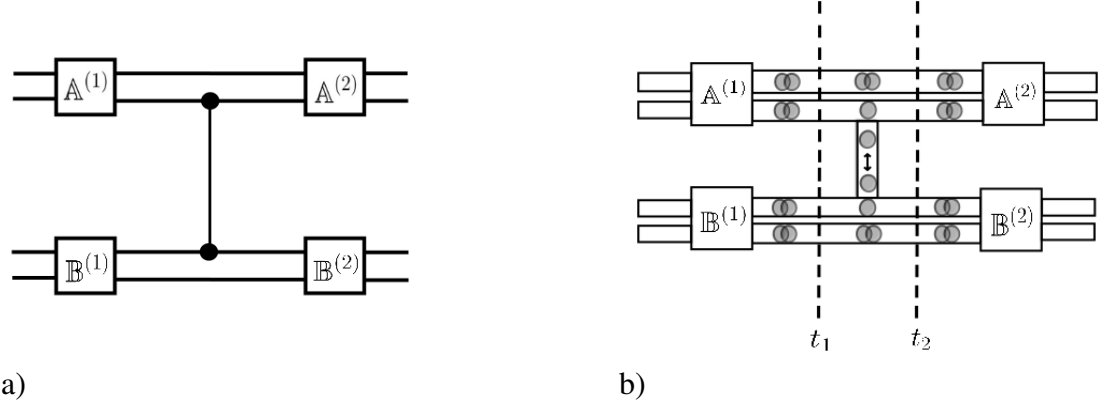

Figure 4: a) Section of a quantum circuit in the dual rail encoding. b) Idealization of this same quantum circuit embedded in three-dimensional space. Circular tokens represent wavepackets propagating through channels formed by their spacetime paths, and are shown at three different times to depict their time evolution. Each channel contains two wavepackets. Alice's lower channel contains a wavepacket corresponding to her part of the state $m_2|1\rangle_A|0\rangle_B$ as well as a wavepacket corresponding to her part of the state $m_4|1\rangle_A|1\rangle_B$. Bob's upper channel contains a wavepacket corresponding to his part of the state $m_1|0\rangle|0\rangle$ and a wavepacket corresponding to his part of the state $m_2|1\rangle_A|0\rangle_B$. An interaction occurs between modes $|1\rangle_A$ of Alice and $|0\rangle_B$ of Bob between times $t_1$ and $t_2$. As represented by the interaction of the two isolated tokens in the vertical channel, we presume that the controlled phase gate only transmits information between Alice's part of the state $m_2|1\rangle_A|0\rangle_B$ and Bob's part of the state $m_2|1\rangle_A|0\rangle_A$. If we make this presumption, particle A cannot produce the marginal probabilities in a local way. To reproduce the marginal probabilities, particle A requires information from particle B during this interaction which is contained in the coefficient $m_1$ of $m_1|0\rangle_A|0\rangle_B$.

## 4.1 Summary of results

This section has formalized a discrepancy between the propagation of information into quantum subsystems and the tensor product structure of the wavefunction. A simple thought experiment provided a proof of this discrepancy. The existence of the strange flow of information into quantum subsystems is difficult to reconcile with Schrödinger time evolution, which contains no provision for the exchange of information beyond that encoded by the tensor product structure of wavefunction. Standard quantum descriptions of interactions, both in the quantum mechanical and quantum field theoretic setting, are potentially limited by this tensor product structure. The main significance of these insights is that any local interpretation of quantum subsystems must account for this peculiar flow of information. Therefore a strong constraint is provided on the possible local interpretations for quantum subsystems.

# 5 Marginals in the sum-over-paths interpretation

The sum-over-paths interpretation describes the quantum predictions in terms of a decomposition of the wavefunction amplitudes into path amplitudes. In the discrete case, the path amplitudes are products of matrix elements of unitary operators. In the continuous case, this interpretation becomes the path integral formulation. We briefly discuss both cases.

## 5.1 Discrete sum-over-paths

Intuitively, it would seem that a sum-over-paths interpretation for marginal probabilities would proceed by summing all path amplitudes of particles of the subsystem to arrive at the detector locations, then taking the modulus squared of the result. However, the sum-over-paths algorithm for the marginals is not so straightforward. The correct sum over paths algorithm consists of the following three steps:

1. Take the sum of all configuration space paths leading to a configuration space point.

2. Take the modulus squared of the result.

3. Do this for all possible positions of particle B, keeping the particular position chosen for particle A fixed. Then take a classical sum over the result.

When the subsystem is entangled to the external system, the particles of the external system cannot be ignored in this procedure. Entanglement causes several problems. First, the marginal probabilities of particle A seem to depend on the final measurement locations of particle B. This information is required to categorize the paths of the subsystem as those which interfere and those which do not. Second, the configuration space paths contain transition amplitudes for particle B to propagate from its last point of interaction with particle A to its measurement location. Removing the dependence on the final measurement location and remnant transition amplitudes of particle B is made difficult by the selective interactions between some of the paths but not others. The selective interaction of the paths of particle B prevents standard methods, such as a no-signaling condition, from removing these dependencies.

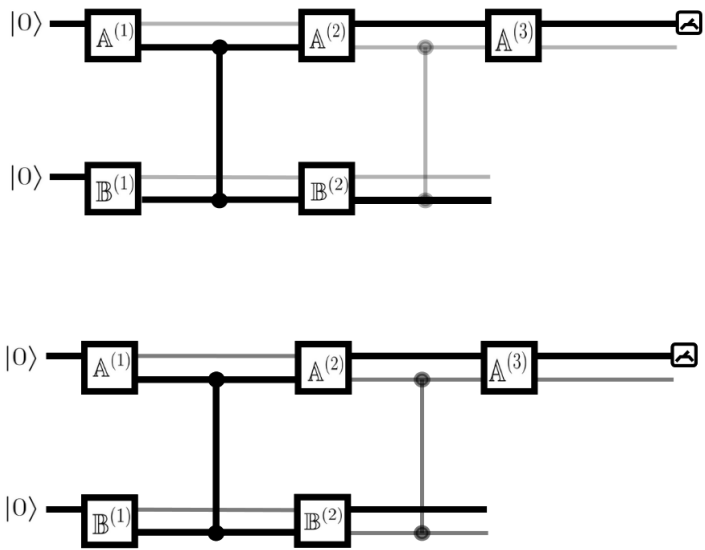

Figure 5: Depiction of two configuration space paths (bolded), superimposed on the quantum circuit of section 2.2 (faded). These two configuration space paths do not interfere to produce the marginal probabilities which Alice observes, despite both of particle A's components of these configuration space paths ending in the same measurement location. Particle A needs to determine whether these two paths interfere quantum mechanically or not, a task which requires knowledge of whether particle B ends at the same measurement location. The standard no-signaling condition cannot be used to remove this dependence on particle B, since there is an A-B interaction occurring later in the circuit (faded).

## 5.2 In the context of information flow hypothesis

One of the main attractions of the sum-over-paths interpretation is that it splits the time evolution of the wavefunction into a set of configuration space paths that can be individually embedded into three-dimensional space. For instance the joint probability density of a bipartite system can be written as $|\Psi(x_1,x_2)|^2 = |\Sigma_r \Sigma_s A_r B_s e^{i\alpha(A_r,B_s)}|^2$ where $A_r B_s e^{i\alpha(A_r,B_s)}$ are amplitudes for the configuration space paths $(r,s)$. We can associate an intuitive three-dimensional space picture to each configuration space path, but only when they are taken in isolation. For example, this picture could be of two particles carrying the values of their single particle amplitudes $A_r$ and $B_s$, along with the phases acquired due to their mutual interaction $e^{i\alpha(A_r,B_s)}$.

A sum-over-paths interpretation of the marginals would be quite elegant if we could regard the individual configuration space paths, when embedded into the ontology for the subsystem, as being independent to one another. However, the sum-over-paths interpretation performs the path decomposition in accordance with the tensor product structure of the wavefunction. Therefore the interpretation is inherently susceptible to the violation of the hypotheses considered in section 4. The violation of these hypotheses indicates that the representations of these configuration space paths in the local interpretation of the subsystem[7] need to communicate information to each other during the application of controlled phase gates.

## 5.3 Path integral formulation

For completeness, we will briefly consider marginals in the path integral formulation. In the path integral formulation, the probability density is given by summing the complex amplitudes $e^{iS_{r,s}}$ of all configuration space paths $(r,s)$ then taking the modulus squared of the result. To produce the marginals, an integral is performed over the positions of particle B, keeping the particular position chosen for particle A fixed. The problems we have outlined for the discrete sum-over-paths are also applicable to the path integral formalism. Although in the continuous case, since each configuration space path has unit weight, there are just non-local phases to contend with. Nevertheless, establishing a local interpretation in the path integral formalism may prove to be even more challenging than for the discrete sum-over-paths. The path integral formalism is not suited to describing discrete pieces of information being carried by paths of the subsystem.

## 5.4 Summary of results

The standard sum-over-paths interpretation describes quantum subsystems in a non-local way. Because the sum is performed over configuration space paths, the marginal probabilities of particle A have a dependence on the final measurement locations and transition amplitudes for particle B. Furthermore, the sum-over-paths interpretation provides a very clear example of the violation of the hypotheses considered in section 4. If there is a one-to-one correspondence between configuration space paths and paths of the subsystem, then the paths of the subsystem need to communicate with each other to enable a local interpretation of the subsystem. These insights are exploited in paper two of this series [12] to establish a local interpretation of quantum subsystems.

---

[7]Assuming some sort of isomorphism between the configuration space paths and the ontological agents of the quantum subsystem (for instance, paths of the subsystem).

# 6  Conclusion

At the quantum level, nature appears remarkably high-dimensional. In the non-relativistic setting, the stage upon which many-body wavefunction dynamics occurs is a 3$N$-dimensional configuration space, not the three-dimensional space in which classical phenomena are grounded. Non-separable superpositions of wavepackets in this configuration space result in correlated joint probabilities for particle measurement that deny a simple local realist explanation. In this paper, we have endeavored to further understand the physical nature of configuration space effects by characterizing their appearance in quantum mechanical marginal probabilities.

This paper demonstrates that the Copenhagen, de Broglie-Bohm and sum-over-paths interpretations in their current form do not provide a local interpretation for quantum marginals. These interpretations ignore the potential locality of the marginals in exchange for describing the joint probabilities in a non-local manner. Such an exchange may be a Faustian bargain, since it is plausible that the marginal probabilities play a central role in the ontology of the full quantum predictions.

However, it is not widely recognized that these interpretations fail to provide a local explanation of the marginals. This fact is obscured by the existence of the no-signaling property, which is often used as proof that the marginals are local. In the case of spatial entanglement, the no-signaling property does not sufficiently remove the configuration space effects arising in the marginal probabilities. We have identified the presence of these configuration space effects in several situations, and have formalized the kernel of intuition underlying the phenomenon. There is a complete dissociation between the tensor product structure of the wavefunction and the local propagation of information into quantum subsystems.

# A  Kronecker product

Note that the Kronecker product of the single particle matrices with the identity can be written as:

$$\mathbb{A}\otimes\mathbb{I} = \begin{pmatrix} A_{00} & 0 & A_{01} & 0 \\ 0 & A_{00} & 0 & A_{01} \\ A_{10} & 0 & A_{11} & 0 \\ 0 & A_{10} & 0 & A_{11} \end{pmatrix} \qquad \mathbb{I}\otimes\mathbb{B} = \begin{pmatrix} B_{00} & B_{01} & 0 & 0 \\ B_{10} & B_{11} & 0 & 0 \\ 0 & 0 & B_{00} & B_{01} \\ 0 & 0 & B_{10} & B_{11} \end{pmatrix}$$

It may be useful to regard the Kronecker product $\mathbb{A}\otimes\mathbb{I}$ as acting the matrix $\mathbb{A}$ on the sub-vectors of the state vector with bases $\{|00\rangle,|10\rangle\}$ and $\{|01\rangle,|11\rangle\}$ respectively. In more physical terms, this means that $\mathbb{A}\otimes\mathbb{I}$ rotates the states of particle A amongst themselves keeping their entanglements with the states of particle B fixed. Similarly $\mathbb{I}\otimes\mathbb{B}$ rotates the states of particle B amongst themselves keeping their entanglements with states of particle A fixed.